\documentclass[prb,twocolumn,floatfix,showpacs,preprintnumbers,groupeaddress,amsmath,amssymb]{revtex4}
\usepackage[dvips]{graphics}

\usepackage{mathptmx}
\usepackage{psfrag}
\usepackage{graphicx}

\begin{document}

\title{Aharonov-Bohm interferometry with quantum dots: scattering approach 
versus tunneling picture}

\author{Bj\"orn Kubala}
\author{J\"urgen K\"onig} 
\affiliation{Institut f\"ur theoretische Festk\"orperphysik, 
Universit\"at Karlsruhe, 
76128 Karlsruhe, Germany}

\date{\today}

\begin{abstract}
We address the question of how to model electron transport through closed 
Aharonov-Bohm interferometers which contain quantum dots.
By explicitly studying interferometers with one and two quantum dots, we 
establish the connection between a tunneling-Hamiltonian formulation on the 
one hand and a scattering-matrix approach on the other hand.
We prove that, under certain circumstances, both approaches are 
{\em equivalent}, i.e., both types of models can describe the {\em same} 
experimental setups.
Furthermore, we analyze how the interplay of the Aharonov-Bohm phase and 
the orbital phase associated with the lengths of the interferometers' arms 
affect transport properties.
\end{abstract}


%
% 73.63.Kv Electronic transport in mesoscopic or nanoscale materials and
%          structures: Quantum dots
% 73.23.Hk Electronic transport in mesoscopic systems:
%          Coulomb blockade; single-electron tunneling
% 73.40.Gk Electronic transport in interface structures: Tunneling
%

\maketitle

\section{Introduction}
The study of transport through mesoscopic multiply connected geometries 
containing one or two quantum dots (QD's) has recently attracted much interest.
For devices smaller than the phase-coherence length, Aharonov-Bohm (AB) 
oscillations, i.e., oscillations of the conductance as a function of magnetic 
flux appear.
This has been experimentally demonstrated for AB interferometers with either 
one QD,\cite{Yacobi95,Schuster97,Ji00,Wiel00,Kobayashi02} or with 2 QD's.\cite{Holleitner00}

\begin{figure}[h]
\centerline{\includegraphics[width=0.85\columnwidth]{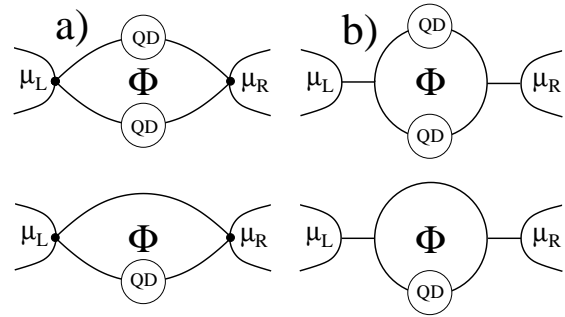}}
\caption{Different AB geometries as described in the text.}
\label{fig0}
\end{figure}

Theoretical discussions of AB interferometry with QD's can be divided into 
two groups.
The first group comprises studies based on a tunnel Hamiltonian approach.
In this case, AB interferometers are modeled as depicted in 
Fig.~\ref{fig0}a): electronic states in the leads are {\em simultaneously} 
coupled to two QD's or to one QD and to the opposite lead.
Using this approach, a variety of phenomena such as resonant 
tunneling\cite{Shahbazyan94,Mourokh} and cotunneling,\cite{Akera93,Loss00} Kondo correlations\cite{Gerland00,Hofstetter01,Boese02,Kim02,Lopez02} 
and Fano physics,\cite{BK,Silva02,Ueda02,Kang02} and the influence of Coulomb interaction on 
quantum coherence\cite{Bruder96,Koenig01,Koenig01.2,Gefen02} has been addressed.
In the second group of papers a scattering-matrix formulation is used.
The AB interferometer is modeled as a ring attached to two leads as shown in 
Fig.~\ref{fig0}b).
After specifying the scattering matrices for the forks connecting the ring
to the leads and for the upper and lower arm, the total transmission through 
the device can be derived.\cite{Buettiker84,Gefen84,Yeyati95,Hackenbroich96,Weidenmueller02,Entin01.2,Aharony02}

The virtue of the tunneling picture lies in the fact that many-body effects,
e.g., due to Coulomb interaction in the QD's, can be included in a conceptually 
straightforward way.
On the other hand, in the scattering-matrix approach the effect of
orbital phases associated with the finite length of the arms of the AB 
interferometer can be discussed in an easy manner.

A question of great importance is how these two types of models are connected 
to each other, and, in particular, to a given experimental setup. 
Do they capture different physical aspects?
At first glance, they seem to describe different geometries, i.e., not to be 
related to each other at all.
In geometries depicted in Fig.~\ref{fig0}b) electrons can travel around the 
ring several times before entering the leads.
In contrast, multiple-loop trajectories around the enclosed flux in 
Fig.~\ref{fig0}a) go through the left and right lead.
One might even argue naively that in the latter case the multiple-loop 
trajectories can not contribute to coherent transport, and thus to 
AB oscillations, since the electrons entering the leads would loose their phase
information immediately.
This notion, however, turns out not to be correct, as we discuss in detail 
below.

To compare the two models we concentrate on situations in which both approaches
are applicable.
As a first condition, we assume the Coulomb interaction among the electrons
in the QD's to be negligible.
In this case, the transmission amplitudes through the QD's, which are essential 
input parameters for the scattering-matrix approach, are easily determined.
(For the problems arising with interacting electrons see the discussion in
Refs.~\onlinecite{Koenig01,Koenig01.2,Gefen02}.)
Secondly, we neglect orbital phases associated with a finite length of the 
interferometers' arms.
We will find that, in complete contrast to the naive expectation formulated 
above, both, the model based on the scattering-matrix formulation and that 
being described by a tunneling Hamiltonian, capture the very same physical 
aspects.
In fact, we explicitly show (for AB interferometers containing one or two QD's)
that both models are even {\em equivalent}, i.e., they yield the same total 
transmission through the device.

This equivalence is even more surprising when realizing that the 
scattering-matrix models seem, at first glance, to have an additional model 
parameter which does not enter the tunneling Hamiltonian.
This parameter is the strength of the coupling between the AB ring and the
leads.
We will resolve this puzzle for the AB interferometers with one or two QD's 
studied in this paper by explicitly deriving scaling relations which 
demonstrate that different ring-lead coupling strengths can be rescaled by
modified tunnel couplings of QD levels.

While the ring-lead coupling strength does not introduce an extra dimension 
for the space of adjustable parameters, the orbital phase associated with the 
finite length of the AB interferometers' arms does. 
In the last part of this paper, we study how the orbital and the AB phase 
affect transport through a double-dot AB interferometer.
We derive an expression for the total transmission under the neglect of 
Coulomb interactions in the QD's.
The result is a generalization of Ref.~\onlinecite{BK}, where, using
a tunneling-Hamiltonian approach, we analyzed the transport features in the 
absence of orbital phases.

\section{Tunnel Hamiltonian versus scattering approach}

\subsection{The tunnel Hamiltonian description for a double-dot Aharonov-Bohm 
interferometer}

We consider an AB interferometer containing two single-level QD's with level
energies $\varepsilon_1$ and $\varepsilon_2$, measured relative to the Fermi 
energy of the leads.
The corresponding tunnel Hamiltonian reads
\begin{equation}
   H = \sum_{kr} \varepsilon_{kr} a^\dagger_{kr}a_{kr} +
   \sum_{i=1,2} \varepsilon_i c^\dagger_i c_i +
   \sum_{kri} (V_{ri} a^\dagger_{k r} c_i + \mathrm{H.c.}) \, ,
\label{Hamiltonian}
\end{equation}
where $a^\dagger_{kr}$ and $a_{kr}$ are the creation and annihilation operators
for electrons with quantum number $k$ in the left or right lead, $r=L$ or $R$,
respectively, and $c^\dagger_i$ and $c_i$ are the Fermi operators for the 
states in dot $i=1,2$.
The AB flux $\Phi$ enclosed by the interferometer arms enters the Hamiltonian
via phase-dependent tunnel matrix elements $V_{ri}$.
In a symmetric gauge the latter can be written as 
$V_{R1} = V_{L2} = |V| \exp (i\varphi/4)$ and 
$V_{R1} = V_{L2} = |V| \exp (-i\varphi/4)$ with 
$\varphi\equiv 2\pi \Phi/(h/e)$.

In our previous work\cite{BK} we have analyzed the transport characteristics 
of the present model in detail.
Here, our goal is the comparison to the scattering-matrix formalism.
Therefore, we just cite the result for the total transmission probability 
through the device for electrons at the Fermi energy
\begin{equation}
  T_\mathrm{tot} = 
  {\Gamma^2 \left[ (\varepsilon_1 + \varepsilon_2)^2 \cos^2 {\varphi \over 2}
      + (\varepsilon_1 - \varepsilon_2)^2 \sin^2 { \varphi \over 2 } \right] 
    \over
    \left[ 2 \varepsilon_1  \varepsilon_2 - {\Gamma^2 \over 2} 
      \sin^2 {\varphi \over 2} \right]^2 + (\varepsilon_1 + \varepsilon_2)^2
    \Gamma^2 }\;,
\label{THamiltonian}
\end{equation}
where the strength of the tunnel coupling of the dot levels to the leads is 
characterized by the intrinsic line width $\Gamma = 2\pi|V|^2(N_L+N_R)$, with 
$N_{L/R}$ being the density of states in the leads at the Fermi energy.

\subsection{The scattering approach for a double-dot Aharonov-Bohm 
interferometer}
\label{scattering}

One-dimensional scattering is a conceptually simple single-particle 
description, whose physical meaning is clear and well understood. 
Scattering matrices ${\mathbf S}_\mathrm{fork}$ for the left and right `fork' 
connecting the AB ring to the left and right lead, respectively, describe how 
an incoming electron from, e.g., the left lead is partially reflected and 
partially transmitted into the upper and lower AB arm.
Transport in the AB arms, which may contain a QD, is modeled by scattering 
matrices ${\mathbf S}_{1/2}$.
Once these scattering matrices are known, the total transmission amplitude 
$t_\mathrm{tot}$ for electrons from the left to the right lead can be easily
obtained by solving a set of linear equations for the appearing amplitudes for 
left- and rightmoving waves in the different parts of the AB interferometer. 

The $3\times 3$ scattering matrix for the fork should satisfy current 
conservation ($ {\mathbf S}_\mathrm{fork}{\mathbf S^{\dagger}_\mathrm{fork}}
={\mathbf S^{\dagger}_\mathrm{fork}}{\mathbf S}_\mathrm{fork}=1$) as well as
 time-reversal symmetry 
(${\mathbf S}_\mathrm{fork}^{-1}={\mathbf S}_\mathrm{fork}^{*}$).
We assume a symmetric geometry of the fork, i.e., symmetry between the two 
channels $1,2$ connections to the upper and lower AB arm.
Furthermore, we restrict ourselves to real entries only. 
With these constraints, the scattering matrix takes either the form\cite{Buettiker84}
\begin{eqnarray}
\label{fork}
{\mathbf S}_\mathrm{fork}(\eta)&=&
        \left( \begin{array}{ccc}
        \eta  & \pm \sqrt{1-\eta^2 \over 2} & \pm \sqrt{1-\eta^2 \over 2}
        \\
        \pm \sqrt{1-\eta^2 \over 2}    & \frac{1-\eta}{2}& -\frac{1+\eta}{2}
        \\
        \pm \sqrt{1-\eta^2 \over 2} &  -\frac{1+\eta}{2} & \frac{1-\eta}{2}
        \end{array} \right)
      \;,
\end{eqnarray}
or ${\mathbf S}'_\mathrm{fork}(\eta)=-{\mathbf S}_\mathrm{fork}(\eta)$.
The first column/row corresponds to the channel coming from/going to the lead, 
and the other two columns/rows are associated with the channels from/to the 
arms of the AB interferometer.
The parameter $\eta$ with $-1 \le \eta \le 1$ characterizes the coupling of
the ring to the leads:
\emph{Strong coupling} of the ring to the leads corresponds to $\eta=0$. 
In this case, no backscattering of a wave into the lead occurs.
\emph{Weak coupling} is realized for $|\eta|\rightarrow 1$.
For $|\eta|=1$, ring and leads are decoupled.
\emph{Symmetric coupling} occurs for $\eta= 1/3$.
In this case, the fork does not distinguish between the lead and the ring 
channels.
For now we use the scattering matrix ${\mathbf S}_\mathrm{fork}(\eta)$ -- 
we will discuss the implications of choosing this branch 
as compared to ${\mathbf S}'_\mathrm{fork}(\eta)$ below.

\begin{figure}
\centerline{\includegraphics[width=0.6\columnwidth]{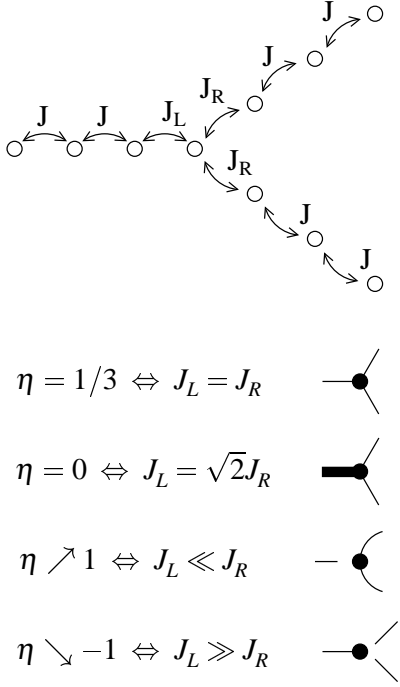}}
\caption{The tight-binding model of a fork allows for a connection of the 
  parameter $\eta$ of the scattering matrix to microscopic parameters: 
  $\; \eta=[2-\left(J_L/J_R\right)^2]/[2+\left(J_L/J_R\right)^2]$. }
\label{figtight1}
\end{figure}

A microscopic understanding of the coupling parameter $\eta$ can be gained
from a tight-binding model of the fork, as shown in Fig.~\ref{figtight1}.
In such a model the fork consists of three semi-infinite chains, with hopping 
matrix elements $-J$ and zero site energies; these leads are connected to a 
central site by couplings $-J_L, -J_R$. 
As shown in Ref.~\onlinecite{Itoh}, the form of Eq.~(\ref{fork}) is 
found for half-filling, i.e. $k_Fa=\pi/2$, where $k_F$ is the Fermi wavevector 
and $a$ the distance of the tight-binding sites.
At this value little energy (or $k$) dependence of the matrix entries was 
found.
Therefore the half-filling case and thus a real scattering matrix can be seen 
as generic for forks with energy-independent scattering. 

The parameter $\eta$ is connected to the microscopic quantities by
$\eta=\left[2-(J_L/J_R)^2\right]/\left[2+(J_L/J_R)^2\right]$.
Thus, the parameter $\eta$ is directly linked to ratio $J_L/J_R$.
In Fig.~\ref{figtight1} we show the respective counterparts of the special 
cases of strong, weak and symmetric coupling in the tight-binding model. 

Next, we construct the scattering matrix for the AB arms.
The transmission and reflection amplitudes for a single-level QD with
level energy $\varepsilon$ and coupling $\Gamma$ to the leads are
\begin{equation}
  t_i=\frac{i \Gamma/2}{-\varepsilon_i+i\Gamma/2}=t^\prime_i
  \;;\quad 
  r_i=\frac{\varepsilon_i}{-\varepsilon_i+i\Gamma/2}=r^\prime_i \;,
\label{TQD}
\end{equation}
where $t_i,r_i$ and $t'_i,r'_i$ are associated with incoming waves from the 
left and right hand side, respectively.
But the scattering matrix should also include phase factors which arise due to 
an AB flux and a finite length of the arms.
For a symmetric geometry, i.e., identical length of all arms, the scattering 
matrices assume the form
\begin{equation}
{\mathbf S}_{1/2} = 
        \left( \begin{array}{ccc}
        r_{1/2}  & t_{1/2}e^{\mp i\varphi/2}
        \\
        t_{1/2} e^{\pm i\varphi/2} & r_{1/2}
        \end{array} \right) \cdot e^{i\beta/2}  \, ,    
\label{ABphase}
\end{equation}
where the first column/row corresponds to channels coming from/going to the 
left side and the second one is associated with the channel on the right hand 
side.
Note that the AB phase $\varphi$ and the orbital phase $\beta$ enter in a 
different manner [$\varphi = 2\pi \Phi/(h/e)$ and $\beta = k_F L$ with the 
Fermi wavevector $k_F$ and the length $L$ of the ring are the AB and the 
orbital phase for a full closed loop around the ring].

To compare with the tunnel-Hamiltonian approach we concentrate on 
$\beta =0$.
In this case one finds after some algebra
\begin{equation}
  T_\mathrm{tot} = 
  {(\alpha_\eta \Gamma)^2 \left[ (\varepsilon_1 + \varepsilon_2)^2 
      \cos^2 {\varphi \over 2}
      + (\varepsilon_1 - \varepsilon_2)^2 \sin^2 { \varphi \over 2 } \right] 
    \over
    \left[ 2 \varepsilon_1  \varepsilon_2 - {(\alpha_\eta \Gamma)^2 \over 2} 
      \sin^2 {\varphi \over 2} \right]^2 + (\varepsilon_1 + \varepsilon_2)^2
    (\alpha_\eta \Gamma)^2 }\;,
\label{general}
\end{equation}
with $2\alpha_\eta=(1+\eta)/(1-\eta)$.
The relation to the tight-binding hopping matrix elements is given by
$\alpha_\eta = (J_R / J_L)^2$.
For the special choice of a symmetric fork, $\eta=1/3$, we find
$\alpha_\eta=1$, and Eq.~(\ref{general}) is
identical to the tunnel Hamiltonian result Eq.~(\ref{THamiltonian}).
This establishes the fact that, although the geometric representations 
Figs.~\ref{fig0}a) and \ref{fig0}b) of the two model systems look different, 
the same physical situation is described.
A different ring-lead coupling, parametrized by $\eta$, does not change
this conclusion.
The only effect is a renormalization of the level broadening 
$\Gamma \rightarrow \alpha_\eta \Gamma$,\footnote{\label{broadening}The reader may have noticed, that the way $\Gamma$ scales with the coupling parameter $\eta$ is rather counterintuitive. A decoupling of ring and leads as $\eta \rightarrow 1$, leads to increased (diverging) level broadening $\Gamma$. Below we will analyze this renormalization in a more general context and will resolve the apparent contradiction.} which can be expressed in the 
scaling relation
\begin{equation}
  T_\mathrm{tot}(\eta,\Gamma,\varepsilon_1,\varepsilon_2,\varphi)
  =
  T_\mathrm{tot}(1/3,\alpha_\eta\Gamma,\varepsilon_1,\varepsilon_2,\varphi)
  \; .
\label{doubledot_renorm}
\end{equation}

%%%%%%%%%%%%%%%%%%%%%%%%%%%%%%%%%%%%%%%%%%%%%%%%%%%%%%%%%%%%%%%%%%%%%%%%%%%%%%%%%%%%%%%%%%%%%%%%%%%%%%%%%%%%%%%%%%%%%

\subsection{Single-dot Aharonov-Bohm interferometer}
\label{singledot}

In the same way as for the double-dot AB interferometer the equivalence of the
tunnel-Hamiltonian and the scattering-matrix approach can be shown for a
single-dot AB interferometer.
The corresponding tunnel Hamiltonian
\begin{eqnarray}
   H &=& \sum_{kr} \varepsilon_{kr} a^\dagger_{kr}a_{kr} +
   \varepsilon c^\dagger c +
   \sum_{kr} (V_{r} a^\dagger_{k r} c + \mathrm{H.c.}) 
\nonumber \\
   && 
   +\sum_{kk'} (W a^\dagger_{kR}  a_{k'L} + \mathrm{H.c.}) 
   \, ,
\label{Hamiltonian_1dot}
\end{eqnarray}
includes, in addition to terms which describes tunneling from the leads to a 
QD and vice versa, a part for direct tunneling from one lead to the other with 
tunnel amplitude $W$.
The strength of the coupling is characterized by the dimensionless parameter
$x = \pi^2 |W|^2 N_L N_R$. 
We model the AB flux using a gauge in which 
$V_L = V_R = |V|$ and $W = |W| e^{i\varphi}$.

A more general model, including a charging energy term for the QD, has been 
studied in Ref.~\onlinecite{Hofstetter01}, and an exact expression for the 
transmission has been derived.
In the absence of charging energy the result can be written as
\begin{equation}
  T_\mathrm{tot} = {\left( 4\sqrt{x} \, \varepsilon/\Gamma-\cos\varphi\right)^2
    +\sin^2 \varphi \over
    \left[ 2(1+x)\varepsilon/\Gamma - \sqrt{x} \,\cos\varphi\right]^2 +1 } \, .
\end{equation}

The analysis of the single-dot AB interferometer within the scattering 
approach is analogous to the case of 2 QD's.
The only difference is that the scattering matrix $\mathbf S_1$ for the
upper arm has to be replaced by
\begin{equation}
{\mathbf S}_\mathrm{dir}=
        \left( \begin{array}{cc}
        -|r|  & i|t| e^{-i\varphi}
        \\
        i|t| e^{i\varphi} & -|r|
        \end{array} \right)\; ,
\label{strong_scatterer}
\end{equation}
where, in order to be able to make connection to the tunnel Hamiltonian
formalism, we modeled the arm by a strong scatterer with $|t| \ll 1$.
It is straightforward to show \cite{Hofstetter01} that
$|t|=2\sqrt{x}/(1+x)$ and $|r|=(1-x)/(1+x)$.

Solving the set of equations for the various amplitudes, we get the total
transmission as
\begin{equation}
  T_\mathrm{tot} = {\left( 4\sqrt{x} \, \varepsilon/\Gamma-\cos\varphi\right)^2
    +\sin^2 \varphi \over
    \left[ 2(1+\alpha^2_\eta x)\varepsilon/(\alpha_\eta\Gamma) 
      - \alpha_\eta\sqrt{x} \, \cos\varphi\right]^2 +1 } \, .
\end{equation}
Again we see that for the special choice $\eta=1/3$, i.e., $\alpha_\eta = 1$, 
for the forks we reach complete equivalence of the tunnel-Hamiltonian and
the scattering-matrix approach.
Different values of $\eta$ can be incorporated as a renormalization of
the coupling constants $\Gamma$ and $x$ for the dot-lead and the direct
lead-lead coupling, respectively.
This leads to the scaling relation
\begin{equation}
  T_\mathrm{tot}(\eta,\Gamma,x,\varepsilon,\varphi) = 
  T_\mathrm{tot}(1/3,\alpha_\eta\Gamma,\alpha_\eta^2 x,\varepsilon,\varphi)
  \, .
\label{singledot_renorm}
\end{equation}

%%%%%%%%%%%%%%%%%%%%%%%%%%%%%%%%%%%%%%%%%%%%%%%%%%%%%%%%%%%%%%%%%%%%%%%%%%%%%%%%%%%%%%%%

\subsection{Connection between the two approaches}

The main conclusion to be drawn from the results derived above is the 
observation of a close connection between the two models based on the 
tunnel-Hamiltonian and the scattering-matrix formalism, respectively.
The two models appear different in the way how the AB interferometer is 
coupled to the the leads, as depicted in Figs.~\ref{fig0}a) and \ref{fig0}b).
It turned out, however, that they can describe the same physical realizations.
To establish this fact, we concentrated on the regime in which both approaches
are valid and easily applied.
We considered noninteracting electrons and small AB interferometers, such that
orbital phases can be neglected.
Furthermore, in the case of the single-dot AB interferometer we assumed small
transparency through the direct arm.

In the scattering approach, the total transmission depends on the strength of
the coupling between the AB ring and the leads, i.e., on the choice of
the scattering matrix for the forks.
We found that the total transmission is identical to the one obtained from the 
tunnel-Hamiltonian approach in the case of symmetric forks, $\eta = 1/3$.
Of course, the ring-lead coupling strength can be different in different
experimental setups.
However, different values of $\eta$ can be modeled by a simple renormalization
of the dot-lead and lead-lead coupling strengths $\Gamma$ and $x$, as 
indicated in the scaling relations Eqs.~(\ref{doubledot_renorm}) and 
(\ref{singledot_renorm}).
If $\Gamma$ and $x$ are viewed as fit-parameters to be determined from the 
experiment, then $\eta$ can be chosen as $1/3$ without loss of generality.

There is another observation which distinguishes the choice $\eta=1/3$ from
other values.
Let us assume that the transmission amplitudes $t_i$ through the arms $i=1,2$ 
are small, $|t_i|,|t_i'|\ll 1$ and $r_i,r'_i \approx -1$.
In this case, all trajectories with higher winding numbers around the AB ring
are negligible, and simply the two paths through either arm 1 or 2 participate.
From the scattering formalism we get the total transmission amplitude
$t_\mathrm{tot} = \alpha_\eta(t_1+t_2)$, which yields the intuitive 
formula for constructive interference $t_\mathrm{tot} = t_1+t_2$ for the
special choice $\eta =1/3$.

This finishes the first part of this paper, which was to demonstrate the 
equivalence between the tunnel-Hamiltonian and the scattering matrix approach.
In the remaining part of this paper we study the effect of the orbital phase,
which we have neglected so far.
We study an AB interferometer with two QD's for noninteracting electrons by 
using the scattering-matrix approach.

\section{The effect of orbital phases on transport}

The finite length of arms of the interferometers can easily be accounted for in
the scattering approach by introducing an orbital phase $\beta$ which enters
the scattering matrices for the two arms, as specified in Eq.~(\ref{ABphase}).
For simplicity, we restrict here to the case of a symmetric sample, i.e., all 
lengths from a fork to a QD are equal.
We first show that also in the presence of the orbital phase $\beta$ models 
with different lead-ring coupling strength $\eta$ can be mapped onto each other
by renormalizing the energy position $\varepsilon_i$ and intrinsic width
$\Gamma$ of the QD levels as well as the orbital phase.
The explicit scaling relations will be derived.
Then, we analyze the transport characteristics and its dependence on the 
orbital phase $\beta$ for $\eta=1/3$.

\subsubsection{Scaling relations for different lead-ring coupling strengths}

\begin{figure}
\centerline{\includegraphics[width=\columnwidth]{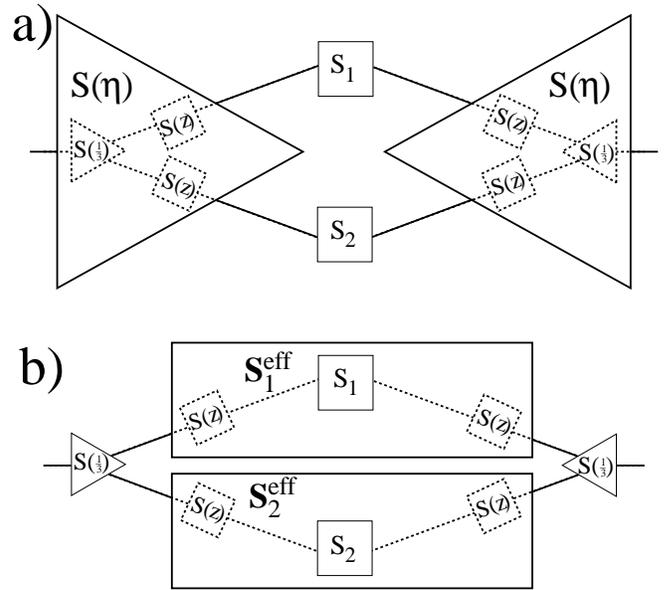}}
\caption{
  a) Disconnection of the scattering matrix of the fork $\mathbf{S}(\eta)$ 
  and b) reconnection to the new scatterers $S_i^\mathrm{eff}$.}
\label{connections}
\end{figure}

We start with the observation that a general fork matrix 
${\mathbf S}_\mathrm{fork}(\eta)$ can be split into a special one (here we 
choose the symmetric fork matrix, $\eta = 1/3$) and two additional scatterers 
in the outgoing interferometer arms.
This is possible, as such a combination of fork matrix and two identical 
scatterers in the outgoing arms (with real entries only) will combine to a 
total scattering matrix fulfilling the constraints of unitarity, time-reversal 
symmetry and real entries, thus be of the general form 
${\mathbf S}_\mathrm{fork}(\eta)$. 
The additional scatterers are then described by  
\begin{equation}
{\mathbf S}={\mathbf S}(z)=\left( \begin{array}{ccc}        
z&\sqrt{1-z^2}      \\
       \sqrt{1-z^2}& -z
       \end{array} \right)\;,
\label{scatterer}
\end{equation} 
with $z= (\eta -1/3)/(1-\eta/3) = (\alpha_\eta - 1)/(\alpha_\eta + 1)$.
These scattering matrices can be combined with the scattering of the AB arms.
We, thus, end up with an effective model containing a symmetric fork, 
$\eta=1/3$, and effective scattering matrices ${\mathbf S}_i^\mathrm{eff}$
instead of ${\mathbf S}_i$ as given in Eq.~(\ref{ABphase}).
The described scheme is visualized in Fig.~\ref{connections}.

\begin{figure}
\centerline{\includegraphics[width=0.9\columnwidth]{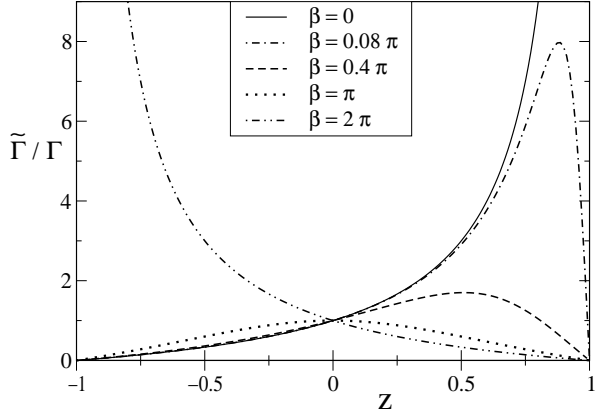}}
\caption{
  Renormalization of the level coupling by additional scatterers $S(z)$ for 
  different geometric phases $\beta$. 
  Note the invariance $\tilde \Gamma(\beta)=\Gamma$ for transparent scatterers
  $z=0$, i.e., $\eta=1/3$.} 
\label{Gammarenorm}
\end{figure}

We find that the effective scattering matrix ${\mathbf S}_i^\mathrm{eff}$ has 
the same structure as ${\mathbf S}_i$ but with renormalized parameters
\begin{subequations}
\label{renrel}
\begin{eqnarray}
  \Gamma &\rightarrow& \tilde \Gamma = 
  \Gamma \frac{1-z^2}{1+z^2-2z\cos{\beta \over 2}}  \, ,
  \label{renrel_Gamma}\\
  \varepsilon_i &\rightarrow& \tilde \varepsilon_i = 
  \varepsilon_i +\Gamma \frac{z\sin{\beta \over 2}}{1+z^2-2z\cos{\beta \over 2}} \, ,
  \label{renrel_eps}\\
  \beta &\rightarrow& \tilde \beta = \beta +2
  \arg{\left(\frac{1-ze^{-i\beta/2}}{1-ze^{i\beta/2}}\right)}\, ,
  \label{renrel_beta} 
\end{eqnarray}
\end{subequations}
i.e., we obtain the scaling relation
\begin{equation}
  T_\mathrm{tot}(\eta,\Gamma,\varepsilon_1,\varepsilon_2,\beta,\varphi)
  =
  T_\mathrm{tot}(1/3,\tilde \Gamma,\tilde \varepsilon_1, \tilde \varepsilon_2,
  \tilde \beta, \varphi)
  \; .
\end{equation}
The renormalization of $\Gamma$ is particularly interesting. 
The level width $\tilde \Gamma$ is reciprocally proportional to the dwell time 
in the scattering region.
It turns out that adding the scatterers ${\mathbf S} (z)$ on the left and 
right hand side of the QD described by ${\mathbf S}_i$ does not necessarily 
increase the dwell time and, therefore, decrease $\tilde \Gamma$.
Due to interference the width can be both increased or decreased, depending on 
the parameters.
This is shown in Fig.~\ref{Gammarenorm}.
We find, that the renormalization of $\tilde\Gamma$ as a function of $z$ shows 
totally different behavior for different values of the orbital phase $\beta$. 
The decoupling limit $\eta \rightarrow 1$, where the ring is separated from 
the leads, corresponds to $z\rightarrow 1$, e.g., a quantum dot 
embedded in between two totally reflecting barriers. 
An electron can not escape from such a structure, thus its dwell time is 
infinite, consequently the resonance width goes to zero. 
Indeed the effective resonance width $\tilde\Gamma$ shows this behavior, 
except for the singular point $\beta=0$, where $\tilde\Gamma$ even diverges 
(see also Sec.~\ref{weak coupling}). 

\subsubsection{Transmission for general orbital phase}

Since we just proved that a system with an arbitrary lead-ring coupling $\eta$ can
be mapped onto an effective model with $\eta = 1/3$, we can restrict to the
latter case for analyzing transport through the AB interferometer.
To calculate the total transmission with an arbitrary orbital phase $\beta$ 
we proceed in the same way as above by setting up a system of linear equations 
for the different amplitudes of the left/right moving waves.
We find for the total transmission the lengthy but complete 
expression
\begin{eqnarray} 
  \label{T_beta}
  t&=&
  16i\Gamma e^{i\beta} 
  \nonumber \\
  && \left[\Gamma \sin{\beta \over 2} \cos{\varphi \over 2} 
      -(1+\cos{\beta \over 2})(\varepsilon_1 e^{i\varphi/2} +
      \varepsilon_2 e^{-i\varphi/2})\right] \nonumber\\
  && \bigg\{ \Gamma^2 \left[16\cos^2{\varphi\over 2}(2\cos{\varphi}-1)
    -(e^{i\beta/2}-1-4\cos{\varphi})^2 \right] \nonumber\\
  && +2i\Gamma(\varepsilon_1+\varepsilon_2)(e^{i\beta}+3)(e^{i\beta/2}+1)(e^{i\beta/2}-3)\nonumber\\
  &&  +4\varepsilon_1\varepsilon_2
  (e^{i\beta/2}+1)^2(e^{i\beta/2}-3)^2   \bigg\}^{(-1)} \, .
\end{eqnarray}
We remark that the transmission probability $T_\mathrm{tot} = |t|^2$ is
symmetric under $\varphi \rightarrow -\varphi$, which is a direct consequence
of Onsager relations and referred to as ``phase locking''.
Furthermore, there is the symmetry
$T_\mathrm{tot}(\beta,\varepsilon_1,\varepsilon_2) = T_\mathrm{tot}
(-\beta,-\varepsilon_1,-\varepsilon_2)$.

\begin{figure}
\includegraphics[width=\columnwidth]{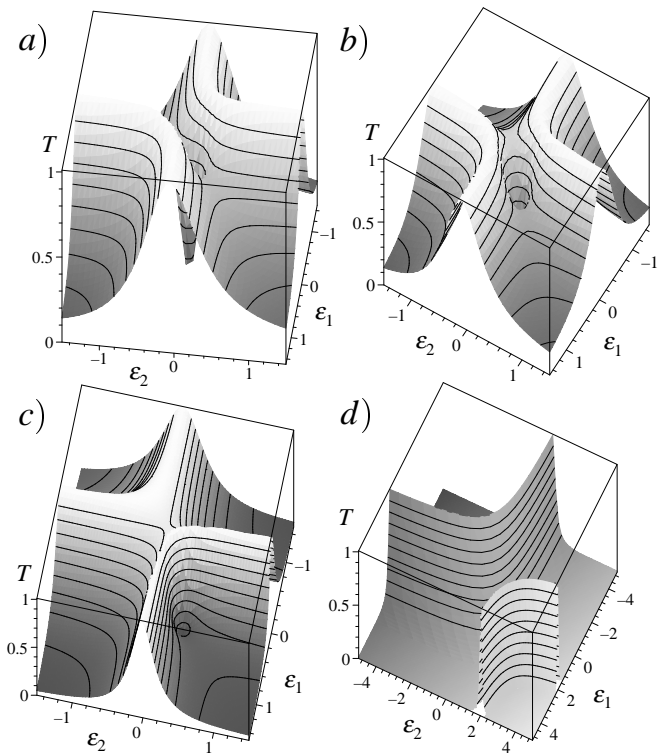}
\caption{
  Transmission $T_\mathrm{tot}=|t|^2$ as given by Eq.~(\ref{T_beta}) as a 
  function of the dots' energies $\varepsilon_1, \varepsilon_2$ for an AB flux 
  $\varphi=0.4\pi$ and different orbital phases: a) $\beta=0$, 
  b) $\beta=0.1\pi$, c) $\beta=2\pi/3$, d) $\beta=1.8\pi$.   
} 
\label{Tlandscape}
\end{figure}

In the absence of an orbital phase, $\beta = 0$, the transmission
probability $T_\mathrm{tot} = |t|^2$ simplifies to Eq.~(\ref{THamiltonian}).
We have analyzed the resulting transport features in Ref.~\onlinecite{BK}.
We found that, at finite flux, the resonances of transmission through the 
total AB interferometer appear at $\varepsilon_1 \varepsilon_2 = - (\Gamma/2)^2
\sin^2(\varphi/2)$, see also Fig.~\ref{Tlandscape}a.
Centered around the region $\varepsilon_1 \approx 0 \approx \varepsilon_2$, 
there is a region of low transmission.
For details see Ref.~\onlinecite{BK}.

We now discuss how a finite orbital phase $\beta$ modifies the transmission
landscape, i.e., the transmission as a function of the QD level energies
$\varepsilon_1$ and $\varepsilon_2$, measured relative to the Fermi energy
of the leads.
As $\beta$ is increased the crater of suppressed transport moves out of the
center along the diagonal $\varepsilon_1=\varepsilon_2$.
Equation~(\ref{T_beta}) yields zero transmission for 
$\varepsilon_1 = \varepsilon_2 = (\Gamma/2)\sin(\beta /2)/[1+\cos(\beta /2)]$.
The two ridges of full transmission merge with increasing $\beta$ and 
eventually split up again, as can be seen in Fig.~\ref{Tlandscape}.
When approaching the limit $\beta \rightarrow 2\pi$, a dramatic change happens.
The transmission peaks become more and more sharp.
In addition, they move along the diagonal $\varepsilon_1=\varepsilon_2$ out
of the center (for $\varphi \ne 0$).

In the limiting case $\beta =2\pi$, the transmission vanishes throughout 
the whole parameter plane $\varepsilon_1,\varepsilon_2$ (for $\varphi\neq 0$).
This somewhat astonishing result can be also understood within a 
tight-binding model, as we carry out in the appendix.

\begin{figure}
\includegraphics[width=0.85\columnwidth]{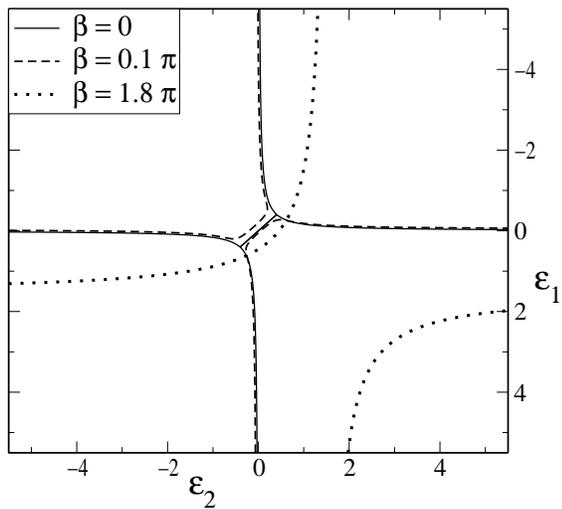}
\caption{
  The resonances of transmission, as given by the zeros of the denominator of 
  $t$ for $\varphi=0.4\pi$. 
  For $\beta\rightarrow 2\pi$ both resonance lines move to $\pm \infty$ (for 
  $\beta\stackrel{<}{\scriptstyle >}2\pi$, respectively).} 
\label{resonances}
\end{figure}

In Fig.~\ref{resonances} we visualize the behavior of the resonance peaks, by 
plotting the zeroes of the denominator of the transmission. 
They are close to the actual resonances $T\equiv1$. 
For both $\beta=0$ and $\beta \rightarrow 2\pi$ the ridges are described by 
hyperbolas, but with different orientation in the 
$\varepsilon_1 - \varepsilon_2$-plane.
For all other values of $\beta$ the ridges are characterized by fourth-order 
curves.

\begin{figure}
\includegraphics[width=0.8\columnwidth]{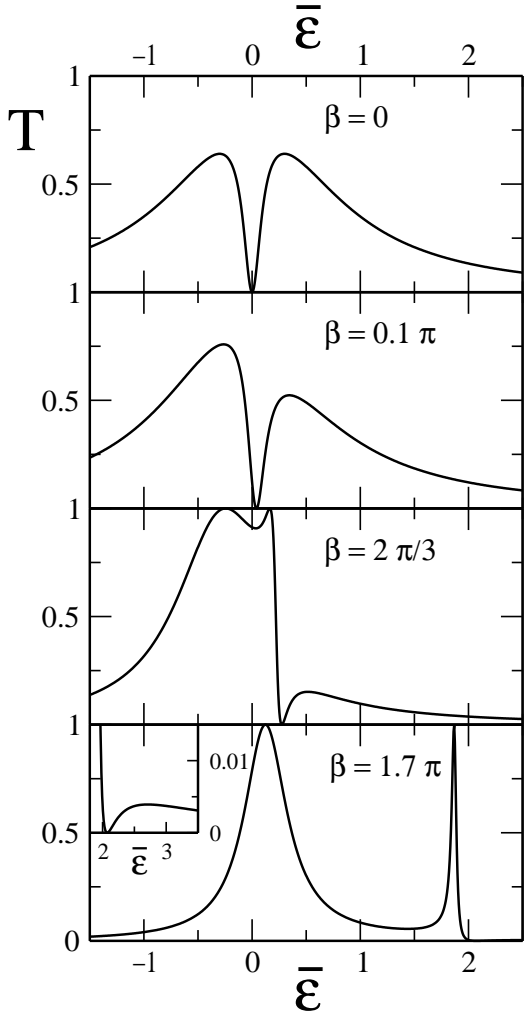}
\caption{
  Transmission along the $\varepsilon_1=\varepsilon_2$ line.
  The value of the ratio 
  $r:=|\varepsilon'_{1}-\varepsilon'_{2}|/(\Gamma'_{1}+\Gamma'_{2})$
  determines the peak structure. 
  We chose $\beta=0, 0.1\pi, 2\pi/3$, and $1.7 \pi$, corresponding to
  $r=0, 0.1, 0.9$, and $6.7$, respectively.}
\label{doublepeak}
\end{figure}

To illustrate the position of the ridges and the crater of suppressed transport
in a more quantitative way, we analyze the transmission along the diagonal 
$\varepsilon_1=\varepsilon_2=\bar\varepsilon$ in more detail.
In this case, the transmission amplitude can be rewritten as
\begin{equation}
  t(\varepsilon_1=\varepsilon_2)=\frac{\Gamma'_{1}}
  {\bar\varepsilon-\varepsilon'_{1}+i\Gamma'_{1}} - 
  \frac{\Gamma'_{2}}{\bar\varepsilon-\varepsilon'_{2}+i\Gamma'_{2}}
\label{T_twolevel}
\end{equation}    
with 
\begin{subequations} 
\label{apex_pos}
\begin{eqnarray} 
  \varepsilon'_{1/2} &=& {1 \over N} \sin{\beta\over2} 
  (1-3\cos{\beta\over2}\pm4\cos{\varphi\over2})
\label{apex_posa}\\
   \Gamma'_{1} &=& {1 \over N} 8\cos^2{\beta\over4} \sin^2{\varphi\over4} 
\label{apex_posb}\\
   \Gamma'_{2} &=& {1 \over N} 8\cos^2{\beta\over4} \cos^2{\varphi\over4} \;,
\label{apex_posc}
\end{eqnarray}
\end{subequations}
where $N =(4\cos{\beta\over2}-3\cos{\beta}+7)/\Gamma$.
Note that Eq.~(\ref{T_twolevel}) equals the transmission through two levels 
$\varepsilon'_{1}, \varepsilon'_{2}$ with widths $\Gamma'_{1}, \Gamma'_{2}$, 
where one tunnel coupling features a relative sign (see 
Ref.~\onlinecite{Silva02} or the $\varphi=\pi/2$ case in Ref.~\onlinecite{BK}).
For $|\varepsilon'_{1}-\varepsilon'_{2}|\ll \Gamma'_{1} + \Gamma'_{2}$ the 
addition of two peaks with opposite sign results in the dip structure around 
$\bar\varepsilon=0$ with $T<1$ at the maxima, see Fig.~\ref{doublepeak}. 
For $|\varepsilon'_{1}-\varepsilon'_{2}|\gg \Gamma'_{1} + \Gamma'_{2}$ we find two separated peaks of different 
widths with unitary transmission.
The relative sign in the tunneling amplitudes results in a Fano-like 
lineshape with a point of zero transmission, see Fig.~\ref{doublepeak}.

\subsubsection{\label{weak coupling}Weak-coupling limit}

Finally, we comment on the limit of a weak coupling between AB ring and leads.
This corresponds to $\eta \rightarrow 1$, i.e., $\alpha_\eta \rightarrow\infty$
or $z\rightarrow 1$.
Recall that for vanishing orbital phase $\beta=0$ the renormalized level width 
$\tilde\Gamma$ diverges, as shown in Fig.~\ref{Gammarenorm}.
This contradicts the intuitive notion that for a ring which is weakly coupled
to the leads the resonance widths should be infinitesimally small as studied 
in detail for general scatterers by B\"uttiker 
(see Ref.~\onlinecite{Buettiker84}).
This apparent contradiction is resolved by the inclusion of a finite
orbital phase.
For $\beta\neq 0$, Eq.~(\ref{renrel_Gamma}) yields $\tilde\Gamma \rightarrow 0$
in the weak-coupling limit, in accordance with the intuitive notion and
Ref.~\onlinecite{Buettiker84}.

Next, we check that the position of resonances match with the eigenstates of
the ring in the weak-coupling limit.
Solving the one-dimensional Schr\"odinger equation for the ring it is easily 
shown that an eigenstates at the Fermi energy exists for 
\begin{equation}
  \bar\varepsilon = - \frac{\Gamma \cos{\beta \over 2} \pm 
    \sqrt{\Gamma^2\cos^2{\varphi\over2}+(\varepsilon_1-\varepsilon_2)^2
      \sin^2{\beta \over 2}}}{2\sin{\beta \over 2}} 
\label{ringresonances}
\end{equation} 
and $\beta\neq 0$.
(For vanishing orbital phase, $\beta=0$, there is always an eigenstate at 
the Fermi energy for arbitrary dot parameters $\varepsilon_1, \varepsilon_2$.)
We checked numerically, that these resonance lines indeed match the 
positions of the ridges in the transmission landscape for $\eta \rightarrow 1$.

\section{Summary}

We addressed the relation between two different approaches of
modeling AB interferometers with QD's, namely, a tunneling-Hamiltonian 
formalism and a scattering-matrix approach.
We proved that, under certain circumstances, both types of models are 
equivalent.
Furthermore, we derived scaling relations which show that AB interferometers
with different coupling strengths between AB ring and leads can be mapped onto
each other.
Moreover, we analyzed how an orbital phase, associated with a finite length of 
the interferometers' arms, affects the transport characteristics.

\begin{acknowledgments}
We acknowledge helpful discussions with Y. Gefen, G. Sch\"on, A. Silva, and Y. Oreg. 
This work was supported by the Deutsche Forschungsgemeinschaft through Graduiertenkolleg 284, the
Emmy-Noether-program, and the Center for Functional Nanostructures.

\end{acknowledgments} 

\appendix*
\section{Total reflection for $\beta=2\pi$ --- an illustration using a
tight-binding model}
\begin{figure}
\centerline{\includegraphics[width=\columnwidth]{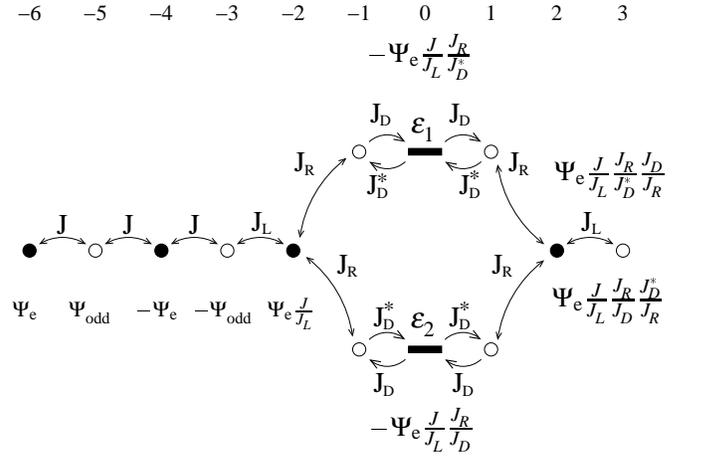}}
\caption{
  In a tight-binding model with odd number of sites in the arms we get 
  expressions for $\Psi_2$ from the upper/lower arm, respectively. 
  This yields $T\equiv0$ for $\varphi\not=0$.}
\label{figtight2}
\end{figure}

In our discussion of the transmission through an AB interferometer as a 
function of the orbital phase $\beta$, we found that for the special case 
$\beta=2\pi$ the total transmission is identical to zero for all values of
the level energies $\varepsilon_1$, $\varepsilon_2$, intrinsic line width 
$\Gamma$, ring-lead coupling $\eta$, and AB phase $\varphi \neq0$.
All incoming electrons will be totally reflected.
In this appendix we show how this behavior can be easily understood by the use 
of a tight-binding model as illustrated in Fig.~\ref{figtight2}.

The left and right leads are represented by semi-infinite chains with hopping
$J$ between adjacent sites, and half filling $k_Fa=\pi/2$.
Coupling of the leads to the AB ring is characterized by hopping matrix 
elements $J_L$ and $J_R$ as discussed in Sec.~\ref{scattering}.
The quantum dot is described by a single site with energy $\varepsilon_i$.
The hopping matrix elements $J_D$ determines $\Gamma$, and the AB phase
$\varphi$ is accounted for by $J_D = |J_D| \exp(i\varphi/4)$.
The orbital phase is due to the finite length of the interferometer. 
At half filling, $k_Fa=\pi/2$, the value $\beta=2\pi$ for the orbital phase 
is achieved by inserting one additional site each between forks and dots, as 
shown in Fig.~\ref{figtight2}.

We remark here that using the alternative choice ${\mathbf S}'_\mathrm{fork}$ 
for the fork scattering matrix introduced in Eq.~(\ref{fork}) together with
$\beta=0$ is equivalent to use ${\mathbf S}_\mathrm{fork}$ and $\beta = 2\pi$.
Thus, adding one site in each arm corresponds to a change between 
${\mathbf S}_\mathrm{fork}$ and ${\mathbf S}'_\mathrm{fork}$ 
(or, equivalently, between $\beta=0$ or $2\pi$), and the transmission through 
the AB interferometer depends on the parity of the number of sites between
dots and forks.

We first look at the solutions of the tight-binding Hamiltonian for an infinite chain 
with one quantum dot embedded.
The solution which describes plane waves far away from the impurity has energy
$E=-2J\cos{k_Fa}$.
For half filling, the corresponding eigenstate $\Psi$ satisfies
\begin{equation}
0=(H-EI)\Psi=  
\left( \begin{array}{ccccc}
        \ddots     &-J & & \cdots &0\\
        -J     & 0  & -J_D& & \vdots \\
               &-J_D&\varepsilon&-J_D\\
         \vdots      &    &-J_D       &   0& -J\\
         0      &  \cdots  &           &  -J&\ddots
        \end{array} \right) \cdot
\left( \begin{array}{c}
\vdots\\
\Psi_{-1}\\
\Psi_0\\
\Psi_{1}\\
\vdots
 \end{array} \right) 
\;.
\label{S_eqn}
\end{equation}      
For $\epsilon=0$ we get two decoupled sets of equations, one for amplitudes on 
sites with an even site number, and one for odd site numbers.
A finite value of $\epsilon$ couples these two sets.

Exploiting this decoupling property, we now calculate the sites' amplitudes $\Psi_i$ in our 
AB setup. 
Assume we know the amplitudes $\Psi_{-6}=:\Psi_{\mathrm e},\;\Psi_{-5}=:
\Psi_{\mathrm odd}\;$ (see Fig.~\ref{figtight2}). 
Using the Schr\"odinger equation (\ref{S_eqn}) for the left chain, we get 
\begin{equation*}
  0=(H\Psi)_{-5}=-J\Psi_{-6}-J\Psi_{-4} \;
  \Rightarrow \Psi_{-4}=-\Psi_{\mathrm e} \; .
\end{equation*}
This procedure is continued along both the upper and the lower arm of the 
interferometer, until we reach the central site of the right fork, whose
amplitude then is
\begin{equation}
  \Psi_{+2}=\Psi_{\mathrm e}\frac{J}{J_L} \frac{J_D}{J^*_D}=
  \Psi_{\mathrm e}\frac{J}{J_L} \frac{J^*_D}{J_D}\; ,
\label{equiv_cond}
\end{equation}
which yields $\Psi_{\mathrm e}=0$ and thus zero transmission $t\equiv 0$ for
all $\varphi$, except for $\varphi=0$.
The full solution, found by using the Schr\"odinger equation for even sites 
(including the dots), delivers a relation linking $\Psi_\mathrm{odd}$ and 
$\Psi_{\mathrm e}$, from which we find $r=-1$.
This explains why for an odd number of sites between dots and forks, i.e., for
$\beta = 2\pi$, the AB interferometer is fully reflecting.

In contrast, in the same geometry but with an even number of sites between QD's and forks, Eq.~(\ref{equiv_cond}) no longer holds.
Instead, the amplitude on the central site of the right fork depends on the 
energies $\varepsilon_1, \varepsilon_2$ of the QD levels, and, in general,
solutions with finite transmission $t$ through the interferometer exist.

In this Appendix we provided an explanation for the surprising result that for 
$\beta=2\pi$ the total transmission is identically zero for all values of
the energies of the QD levels (provided that $\varphi \not=0$).

%%%%%%%%%%%%%%%%%%%%%%%%%%%%%%%%%%%%%%%%%%%%%%%%%%%%%%%%%%%%%%%%%%%%%%%%%%%%%%%%%%%%%


\begin{thebibliography}{99}

\bibitem{Yacobi95}
A. Yacoby, M. Heiblum, D. Mahalu, and H. Shtrikman, Phys. Rev. Lett. {\bf 74},
4047 (1995).

\bibitem{Schuster97}
R. Schuster, E. Buks, M. Heiblum, D. Mahalu, V. Umansky, and H. Shtrikman, Nature (London) {\bf 385}, 417 (1997).

\bibitem{Ji00}
Y. Ji, M. Heiblum, D. Sprinzak, D. Mahalu, and H. Shtrikman, Science {\bf 290}, 779 (2000);
Y. Ji, M. Heiblum, and H. Shtrikman, Phys. Rev. Lett. {\bf 88}, 076601 (2002).

\bibitem{Wiel00}
W.G. van der Wiel, S. De Franceschi, T. Fujisawa, J.M. Elzerman, S. Tarucha, and L.P. Kouwenhoven, Science {\bf 289}, 2105 (2000).

\bibitem{Kobayashi02}
K. Kobayashi, H. Aikawa, S. Katsumoto, and Y. Iye, Phys. Rev. Lett. {\bf 88}, 256806 (2002).

\bibitem{Holleitner00}
A.W. Holleitner, C.R. Decker, H. Qin, K. Eberl, and R.H. Blick, Phys. Rev. 
Lett. {\bf 87}, 256802 (2001).

\bibitem{Shahbazyan94}
T.V. Shahbazyan and M.E. Raikh, Phys. Rev. B {\bf 49}, 17123 (1994).

\bibitem{Mourokh}
L.G. Mourokh, N.J.M. Horing, and A.Yu. Smirnov, Phys. Rev. B {\bf 66}, 085332 (2002).
 
\bibitem{Akera93}
H. Akera, Phys. Rev. B {\bf 47}, 6835 (1993).

\bibitem{Loss00}
D. Loss and E.V. Sukhorukov, Phys. Rev. Lett. {\bf 84}, 1035 (2000).

\bibitem{Gerland00}
U. Gerland, J. v. Delft, T.A. Costi, and Y. Oreg, Phys. Rev. Lett. {\bf 84},
3710 (2000).

\bibitem{Hofstetter01}
W. Hofstetter, J. K\"onig, and H. Schoeller, Phys. Rev. Lett. {\bf 87}, 156803
(2001).

\bibitem{Boese02}
D. Boese, W. Hofstetter, and H. Schoeller, Phys. Rev. B {\bf 66}, 125315 (2002).

\bibitem{Kim02}
T. Kim and S. Hershfield, Phys. Rev. Lett. {\bf 88}, 136601 (2002).

\bibitem{Lopez02}
R. L\'opez, R. Aguado, and G. Platero, Phys. Rev. Lett. {\bf 89}, 136802 (2002).

\bibitem{BK}
B. Kubala and J. K\"onig, Phys. Rev. B {\bf 65}, 245301 (2002).

\bibitem{Silva02}
A. Silva, Y. Oreg, and Y. Gefen, Phys. Rev. B {\bf 66}, 195316 (2002).
 
\bibitem{Ueda02}
A. Ueda, I. Baba, K. Suzuki, and M. Eto, J. Phys. Soc. Jpn. 72, Suppl. A 157 (2003).
 
\bibitem{Kang02}
K. Kang and S.Y. Cho, cond-mat/0210009 (unpublished).

\bibitem{Bruder96}
C. Bruder, R. Fazio, and H. Schoeller, Phys. Rev. Lett. {\bf 76}, 114 (1996).
 
\bibitem{Koenig01}
J. K\"onig and Y. Gefen, Phys. Rev. Lett. {\bf 86}, 3855 (2001).

\bibitem{Koenig01.2}
J. K\"onig and Y. Gefen, Phys. Rev. B {\bf 65}, 045316 (2002).

\bibitem{Gefen02}
Y. Gefen, cond-mat/0207440 (unpublished).

\bibitem{Gefen84}
Y. Gefen, Y. Imry, and M.Ya. Azbel, Phys. Rev. Lett. {\bf 52}, 129 (1984).

\bibitem{Buettiker84}
M. B\"uttiker, Y. Imry, and M.Ya. Azbel, Phys. Rev. A {\bf 30}, 1982 (1984).

\bibitem{Yeyati95}
A. Levy Yeyati and M. B\"uttiker, Phys. Rev. B {\bf 52}, R14360 (1995).

\bibitem{Hackenbroich96}
G. Hackenbroich and H.A. Weidenm\"uller, Phys. Rev. Lett. {\bf 76}, 110 (1996).

\bibitem{Entin01.2}
O. Entin-Wohlman, A. Aharony, Y. Imry, and Y. Levinson, J. Low Temp. Phys. {\bf 126}, 1251 (2002).

\bibitem{Weidenmueller02}
H.A. Weidenm\"uller, Phys. Rev. B {\bf 65}, 245322 (2002).

\bibitem{Aharony02} 
A. Aharony, O. Entin-Wohlman, B.I. Halperin, and Y. Imry, Phys. Rev. B {\bf 66}, 115311 (2002).

\bibitem{Itoh}
T. Itoh, Phys. Rev. B {\bf 52}, 1508 (1995).


\end{thebibliography}
\end{document}